\documentclass[prl,superscriptaddress,twocolumn, nobibnotes]{revtex4}

\usepackage{enumerate}
\usepackage{amsfonts,amssymb,amsmath}
\usepackage[]{graphics,graphicx,epsfig}
\usepackage{amsthm}

\def\identity{\leavevmode\hbox{\small1\kern-3.8pt\normalsize1}}

\usepackage{graphicx}
\usepackage{dcolumn}
\usepackage{natbib}
\usepackage{color}
\usepackage{multirow}

\newcommand{\Mod}[1]{\ (\mathrm{mod}\ #1)}

\begin{document}

%%%%%%%%%%%%%%%%%%%%%%%%%%%%%%%%%%%%%%%%%%%%%%%%%%%%%%%%%%

\title{Disproving hidden variable models with spin magnitude conservation}

\author{Pawe{\l} Kurzy{\'n}ski} 

\affiliation{Faculty of Physics, Adam Mickiewicz University, Umultowska 85, 61-614 Pozna\'n, Poland}

\author{Wies{\l}aw Laskowski}

\email{wieslaw.laskowski@ug.edu.pl}

\affiliation{Institute of Theoretical Physics and Astrophysics, Faculty of Mathematics, Physics and Informatics, University of Gda\'nsk, 80-308 Gda\'nsk, Poland}

\author{Adrian Ko{\l}odziejski}

\affiliation{Institute of Theoretical Physics and Astrophysics, Faculty of Mathematics, Physics and Informatics, University of Gda\'nsk, 80-308 Gda\'nsk, Poland}

\author{K\'aroly F. P\'al}
\address{Institute for Nuclear Research, Hungarian Academy of Sciences, H-4001 Debrecen, P.O. Box 51, Hungary}

\author{Junghee Ryu}
\address{Centre for Quantum Technologies, National University of Singapore, 3 Science Drive 2, 117543 Singapore, Singapore}

\author{Tam\'as V\'ertesi}
\address{Institute for Nuclear Research, Hungarian Academy of Sciences, H-4001 Debrecen, P.O. Box 51, Hungary}

%\date{\today}

\begin{abstract} 
The squares of the three components of the spin-s operators sum up to $s(s+1)$. However, a similar relation is rarely satisfied by the set of possible spin projections onto mutually orthogonal directions. This has fundamental consequences if one tries to construct a hidden variable (HV) theory describing measurements of spin projections. We propose a test of local HV-models in which spin magnitudes are conserved. These additional constraints imply that the  corresponding inequalities are violated within quantum theory by  larger classes of correlations than in the case of standard Bell inequalities. We conclude that in any HV-theory pertaining to measurements on a spin one can find situations in which either HV-assignments do not represent a physical reality of a spin vector, but rather provide a deterministic algorithm for prediction of the measurement outcomes, or HV-assignments represent a physical reality, but the spin cannot be considered as a vector of fixed length.
	\end{abstract}

\maketitle

%%%%%%%%%%%%%%%%%%%%%%%%%%%%%%%%%%%%%%%%%%%%%%%%%%%%%%%%%%

\section{Introduction}

Quantum theory allows to probabilistically predict measurement outcomes, however it says nothing about the underlying physical reality of a system that is measured. In the early days of quantum physics it was often assumed that an extension of the theory using some additional hidden variables (HVs) would provide one with mathematical tools to describe the underlying reality. However, this assumption lead to fundamental physical consequences, such as nonlocality or contextuality \cite{bell1, bell2, leggett, kochen,klyachko,markiewicz,cabello, kurzynski1, kurzynski2}.

In this work we revisit the above problem and show that any HV theory of spin either describes the underlying physical reality, but the magnitude of the spin is not conserved, or the magnitude is conserved, but HVs do not determine the underlying physical reality -- they only provide a deterministic algorithm for calculating the outcomes of measurements. This observation is a nonclassical aspect of spin systems that is related, but distinct, from the Bell~\cite{bell1}, the Kochen-Specker~\cite{kochen}, and the Leggett theorems~\cite{leggett}.

We argue that a physical HV model should not only be local, realistic and noncontextual, but should also obey the conservation laws.
In case of spin systems the model should conserve angular momentum, i.e., the length of spin vectors should be fixed (see \cite{NAGATA, DOWLING, Pitowsky} for another additional constraints on HV models). As a result, our model extends the nonclassical behaviour to a broader class of quantum states. There are spin states that are local and noncontextual, but whose spin vector cannot be described by a magnitude conserving model. Apart from foundational importance, the spin magnitude conserving HV model can have potential applications in various fields of quantum physics. Moreover, we show that the model can be tested in experiments analogous to the ones testing standard locality and noncontextuality. New experiments can be performed and some old ones can be re-examined.

\section{Results}

\subsection{The problem of spin magnitude in hidden variable theories}

 Let us first recall some basic facts. Consider a spin $s$ particle whose spin state is represented by a vector or a density matrix in a $(2s+1)$-dimensional Hilbert space. The number $s$ is either an integer or a half-integer. The average values of spin coordinates can be determined with a help of the three spin operators $\hat{S}_x$, $\hat{S}_y$, and $\hat{S}_z$ \cite{book3}. The spectrum of each operator is: $s,s-1,\ldots,-s$. These eigenvalues correspond to possible spin projections onto a given axis. The spin values are given in the $\hbar$ units and, as is done commonly in the literature, we assume $\hbar=1$. Interestingly, $\hat{S}_x^2+\hat{S}_y^2+\hat{S}_z^2=s(s+1)\hat{\openone}$, where $\hat{\openone}$ is the identity operator. The above formula implies that any spin $s$ state is an eigenvector of the sum of squared spin operators, therefore the value of the spin length is state independent and equal to $\sqrt{s(s+1)}$. In addition, the relation for the sum of squares does not depend on the choice of directions $x$, $y$, and $z$. What matters, is that the directions are mutually orthogonal.

However, imagine that there is a HV theory that assigns a well defined value to each spin operator. More precisely, each spin operator $\hat{S}_n$, where $n$ denotes the direction in space, is preassigned one of $2s+1$ eigenvalues, say $v(\hat{S}_n)=s-1$. The Kochen-Specker theorem \cite{kochen} states that for $s \geq 1$ such assignment is not possible for some properly chosen sets of directions, but it is in principle possible to do it for three mutually orthogonal directions $x$, $y$ and $z$ for any $s$.

Here, we focus on the following problem. We know that it is possible to deterministically assign eigenvalues to three spin operators for  mutually orthogonal axes. This results in a HV description of spin vector ${\mathbf s}=(v(\hat{S}_x),v(\hat{S}_y),v(\hat{S}_z))$. However, is it possible to satisfy ${\mathbf s}\cdot {\mathbf s}=v(\hat{S}_x)^2+v(\hat{S}_y)^2+v(\hat{S}_z)^2=s(s+1)$?

The above question has a positive answer for some values of $s$, but before we provide a general answer, let us consider three examples. For $s=1/2$ the HV model leads to ${\mathbf s}=(\pm \frac{1}{2},\pm \frac{1}{2},\pm \frac{1}{2})$, hence
${\mathbf s}\cdot {\mathbf s}=3/4=s(s+1).$ Therefore, for $s=1/2$ a HV model naturally conserves the spin magnitude.

For $s=1$ one has ${\mathbf s}\cdot {\mathbf s} = s(s+1)=2$ and the spin magnitude is conserved for HV assignments of the form $(\pm 1,\pm 1,0)$, $(\pm 1,0,\pm 1)$, or $(0,\pm 1,\pm 1)$. Therefore, it is possible to construct a spin magnitude-preserving HV model by carefully choosing proper eigenvalue assignments.

The smallest $s$ for which all HV assignments do not satisfy ${\mathbf s}\cdot {\mathbf s} = s(s+1)$ is $s=3/2$. This time the eigenvalues of a spin operator are $\pm \frac{3}{2}$ and $\pm \frac{1}{2}$, whereas $s(s+1)=\frac{15}{4}$. All HV assignments can be split into four classes: $(\pm\frac{3}{2},\pm\frac{3}{2},\pm\frac{3}{2})$, $(\pm\frac{3}{2},\pm\frac{3}{2},\pm\frac{1}{2})$, $(\pm\frac{3}{2},\pm\frac{1}{2},\pm\frac{1}{2})$, $(\pm\frac{1}{2},\pm\frac{1}{2},\pm\frac{1}{2})$. They lead to the following values of ${\mathbf s}\cdot {\mathbf s}$: $\frac{27}{4}$, $\frac{19}{4}$, $\frac{11}{4}$ and $\frac{3}{4}$, respectively. None of them is equal to $\frac{15}{4}$, therefore the physical interpretation of the HV model is either that ${\mathbf{s}}$ represents the physical reality but the spin magnitude is not conserved, or that ${\mathbf{s}}$ does not represent the physical reality but provides a deterministic method for calculating outcomes of measurements.

For an arbitrary $s$ the spin magnitude conserving HV model can be in principle constructed for half of the half-integer cases and most of the integer ones (see Methods). For the values of $s$ for which it cannot be constructed one can speak of a phenomenon that is analogous to the state-independent contextuality~\cite{yu, kleinmann}, namely there is no model for any state. For the remaining cases one needs to look for less straightforward methods to refute this model.

Firstly, note that for most $s$ values, even if there is a model, only a limited number of possible spin projections can be measured if the spin magnitude is to be conserved. The remaining projections can never be measured according to the model, so if they are measured, the model is contradicted. For example, for $s=2$ we have $s(s+1)=6$. The only HV assignments allowed by our model are of the form $(\pm 2,\pm 1, \pm 1)$ and all permutations thereof. Nevertheless, it is in principle possible to measure that a projection of the spin onto some axis is zero. This value does not occur in any allowable HV assignment, therefore by measuring projection zero one automatically refutes the model. However, there exist states, such as $|s=2,s_z=1\rangle$, for which the probability of measuring a projection zero along the directions $x$, $y$ or $z$ is zero. This is because this state is orthogonal to $|s=2,s_x=0\rangle$, $|s=2,s_y=0\rangle$, and $|s=2,s_z=0\rangle$. Therefore, the above example can be considered as a state-dependent scenario (with respect to measurements along the $x$, $y$, and $z$ axes).

\subsection{Bell-like scenario}

 In general, the state-dependent HV scenarios are tested with the help of Bell-like inequalities \cite{clauser}. Therefore, our next goal is to find such an inequality to test our spin magnitude conserving model.

Since we are interested in the magnitudes of spin vectors, we should focus on spin measurements along three mutually orthogonal directions $\hat{S}_x$, $\hat{S}_y$, and $\hat{S}_z$. Although these measurements cannot be performed jointly, due to a lack of commutation, they are deeply related because of the physical properties of the model studied by us. However, these three measurements are not enough to construct any meaningful Bell-like inequality. We need more measurements. A natural choice, the one we follow, is to consider a bipartite scenario with a spatially separated pair of spins, which we label A and B, and a set of six measurements:
\begin{eqnarray}
\hat{S}_i^{(A)} &\equiv& \hat{S}_i \otimes \openone, \\
\hat{S}_j^{(B)} &\equiv& \openone \otimes \hat{S}_j,
\end{eqnarray}
where $i,j=x,y,z$.

The general form of the correlation inequality for two spin-$s$ particles and the six spin operators is of the form
\begin{equation}\label{giq1}
c_{xx} \langle \hat{S}^{(A)}_x \hat{S}^{(B)}_x \rangle + c_{xy} \langle \hat{S}^{(A)}_x  \hat{S}^{(B)}_y \rangle + \dots + c_{zz} \langle \hat{S}^{(A)}_z \hat{S}^{(B)}_z \rangle \geq \beta,
\end{equation}
where $c_{kl}$ are real coefficients and $\beta$ is the classical bound derived from our model. The above can be rewritten as
\begin{equation}
\langle \mathbf{S^{(A)}} \cdot \mathbf{C} \cdot \mathbf{S^{(B)}} \rangle \geq \beta,
\label{inequality}
\end{equation}
where $\mathbf{S^{(X)}}=(\hat{S}^{(X)}_x,\hat{S}^{(X)}_y,\hat{S}^{(X)}_z)$ is the vector of operators and
\begin{equation}\label{C}
\mathbf{C}=\begin{pmatrix}
c_{xx} & c_{xy} & c_{xz} \\
c_{yx} & c_{yy} & c_{yz} \\
c_{zx} & c_{zy} & c_{zz}
\end{pmatrix}.
\end{equation}

Before we derive an inequality for general $s$, let us consider few examples for $s=1$ and $s=2$.

Example 1. Consider two spin-1 particles and the following inequality
\begin{eqnarray}
&-& \langle \hat{S}^{(A)}_x \hat{S}^{(B)}_x \rangle - \langle \hat{S}^{(A)}_x \hat{S}^{(B)}_y \rangle  -  \langle \hat{S}^{(A)}_y \hat{S}^{(B)}_x \rangle \\
&+& \langle \hat{S}^{(A)}_y \hat{S}^{(B)}_y \rangle - \langle \hat{S}^{(A)}_z \hat{S}^{(B)}_z \rangle ~~ \ge \beta. \nonumber
\end{eqnarray}
One can easily show by considering all possible HV assignments, which in our model are of the form ${\mathbf{s}}_{X}=(\pm 1,\pm 1,0)$, including all possible permutations, that the classical HV bound taking into account the conservation of the spin vector magnitude is $\beta = -2$.

On the other hand, the classical HV bound which does not take into account the spin magnitude and allows for assignments such as $(\pm 1,\pm 1, \pm 1)$, or $(\pm 1, 0, 0)$, leads to the bound $\bar{\beta}=-3$.

Finally, one can find that in quantum theory the spin magnitude conserving bound can be violated down to $\beta_Q = -\frac{(1+\sqrt{17})}{2}=-2.5616$. 
%%%%%%%%%%%%%%%%%%%%%%%%%%%%%%%%%%%%
This value corresponds to the lowest eigenvalue of the Bell operator $\mathbf{S^{(A)}} \cdot \mathbf{C} \cdot \mathbf{S^{(B)}}$, whereas the corresponding eigenvector is the quantum state that violates the inequality the most. 
%%%%%%%%%%%%%%%%%%%%%%%%%%%%%%%%%%%%
 This means that in the corresponding scenario quantum theory cannot be described by a model in which spin magnitudes are conserved, but can be described, in principle, by the model in which the magnitudes are not conserved. Interestingly, the quantum state giving the maximal violation $\beta_Q$ above is not a maximally entangled two-qutrit state. It is a partially entangled state with two equal Schmidt coefficients, and the third Schmidt coefficient being nonzero. A similar situation occurs in the case of the well-known Collins-Gisin-Linden-Massar-Popescu (CGLMP) inequality \cite{CGLMP}.

Example 2. Consider another inequality and two spin-1 particles
\begin{eqnarray}
&-& \langle \hat{S}^{(A)}_x \hat{S}^{(B)}_x \rangle - \langle \hat{S}^{(A)}_x \hat{S}^{(B)}_y \rangle  - \langle \hat{S}^{(A)}_x \hat{S}^{(B)}_z \rangle  \\
&-&  \langle \hat{S}^{(A)}_y \hat{S}^{(B)}_x \rangle - \langle \hat{S}^{(A)}_y \hat{S}^{(B)}_y \rangle - \langle \hat{S}^{(A)}_y \hat{S}^{(B)}_z \rangle \nonumber \\
&-& \langle \hat{S}^{(A)}_z \hat{S}^{(B)}_x \rangle - \langle \hat{S}^{(A)}_z \hat{S}^{(B)}_y \rangle + 3 \langle \hat{S}^{(A)}_z \hat{S}^{(B)}_z \rangle ~~ \ge \beta \nonumber
\end{eqnarray}
This time $\beta = -4$, $\bar{\beta} = -7$ and $\beta_Q = -\sqrt{17}=-4.1231$.

Example 3. Finally, consider two spin-2 particles and the following inequality
\begin{eqnarray}
& & (5/2)\langle \hat{S}^{(A)}_x \hat{S}^{(B)}_x \rangle +2 \langle \hat{S}^{(A)}_x \hat{S}^{(B)}_y \rangle  - \langle \hat{S}^{(A)}_x \hat{S}^{(B)}_z \rangle  \\
&+& (5/2)\langle \hat{S}^{(A)}_y \hat{S}^{(B)}_x \rangle - 2\langle \hat{S}^{(A)}_y \hat{S}^{(B)}_y \rangle - \langle \hat{S}^{(A)}_y \hat{S}^{(B)}_z \rangle \nonumber \\
&-& (3/2)\langle \hat{S}^{(A)}_z \hat{S}^{(B)}_x \rangle - 3 \langle \hat{S}^{(A)}_z \hat{S}^{(B)}_z \rangle ~~ \ge \beta \nonumber
\end{eqnarray}
For this example $\beta=-20$, $\bar{\beta}=-34$ and $\beta_Q=-20.1897$.

Note that there is a geometric interpretation of the inequalities above. Namely, the set of local distributions with conserved spin magnitude form a polytope $\mathcal{P}$ in the (9-dimensional) space of correlations. This polytope $\mathcal{P}$ is completely characterized by a finite number of vertices, each of them corresponding to a possible deterministic HV assignment. In the HV model, any correlation can be expressed as a convex combination of deterministic HV assigments, hence it is a point lying inside the polytope $\mathcal{P}$. Conversely, any point outside the polytope $\mathcal{P}$ can be detected with inequalities which define the facets of the polytope $\mathcal{P}$. Each of the three examples above are such kind of inequalities, which detect quantum correlations that admit a standard HV model, however, not reproducible with HV models, where the spin magnitude is conserved.

\subsection{Inequality for general $s$}

The inequalities in the above examples seem not to be related and one may get an impression that for each $s$ one needs to derive an independent inequality, which would make the problem much more complicated. However, we show that a universal approach is possible and one can derive an inequality which can be violated for any $s$ (see Methods). The main idea behind the derivation is to use the symmetry of the two spin-$s$ singlet state and to assume that the matrix $\mathbf{C}$ is an orthogonal rotation matrix, i.e., the columns and the rows of $\mathbf{C}$ are normalized and orthogonal and its determinant is one.

\subsection{Example of an inequality for a general $s$}

Consider the following rotation matrix
\begin{equation}
\mathbf{C}=\begin{pmatrix}
\frac{1}{\sqrt{2}} & -\frac{1}{\sqrt{2}} & 0 \\
\frac{1}{\sqrt{2}} & \frac{1}{\sqrt{2}} & 0 \\
0 & 0 & 1
\end{pmatrix}.
\label{pmatrix}
\end{equation}
This matrix leads to a Bell-like inequality (\ref{inequality}) that tests the spin magnitude conserving HV model for two spin-$s$ particles. The inequality is violated by the locally rotated singlet state $|\phi\rangle = \hat U^{(B)}|\psi_0\rangle$, where
\begin{equation}
|\psi_0\rangle = \frac{1}{\sqrt{2s+1}}\sum_{m=-s}^s (-1)^{s-m}|m\rangle \otimes |-m\rangle,
\end{equation} 
and $\hat U^{(B)}$ is the unitary transformation corresponding to rotation of spin $B$ that is generated by the rotation matrix  (\ref{pmatrix}) -- for details see Methods. The quantum value is $-s(s+1)$, which is always less than the classical magnitude conserving bound. In Table~\ref{t_ineq} we present some examples of quantum values and classical bounds.

\begin{table}
	\begin{tabular}{ c c c c }
		\hline \hline
		\\[-1.8ex]
		$s$  & $\beta$ and $\bar \beta$ & $-s(s+1)$  \\
		\hline
		1 & $-1-\frac{1}{\sqrt{2}} \approx -1.707$  & -2 \\
		& $-1-\sqrt{2} \approx -2.414$            &    \\
		\hline
		2 & $-1+\frac{1}{\sqrt{2}}-4 \sqrt{2} \approx$-5.949  & -6  \\
		& $-4 - 4\sqrt{2} \approx -9.657$                   &     \\
		\hline
		3 & $-4(1+\sqrt{2}) \approx -9.657$  & -12 \\
		&  $-9 - 9\sqrt{2} \approx -21.730$ &   \\
		\hline
		4 & $-14\sqrt{2}\approx -19.799$ & -20 \\
		& $-16 - 16\sqrt{2} \approx -38.627 $    &     \\
		\hline \hline
	\end{tabular}
	\caption{Quantum violations of the inequality (\ref{inequality}) with $\mathbf{C}$ given by (\ref{pmatrix}). $\beta$ denotes the classical HV bound taking into account the conservation of the spin vector magnitude, whereas $\bar \beta$ denotes the HV bound without that additional constraint. \label{t_ineq}}
\end{table}

\section{Discussion}

We have shown that if, except for the assumptions of locality, noncontextuality and realism, one also requires that spin magnitudes are conserved in a two spin-$s$ correlation-type experiment, it leads to a stronger version of Bell-Kochen-Specker theorem. Depending on $s$, our findings can be formulated either as a state-independent no-go theorem, or as state-dependent inequalities. The bounds on our inequalities can be derived using either standard local HV (LHV), which do not assume that sum of squares of spin observables is $s(s+1)$, or using our LHV model (call it LHV'), which assumes that such sum gives $s(s+1)$. The bounds derived from LHV' are tighter, therefore the LHV' polytope lies inside the LHV polytope. In other words, any correlation lying inside the LHV' polytope automatically lies inside the LHV one, but there are correlations inside the LHV polytope, but outside the LHV' one. Hence, the inequalities with the additional constraints are violated by a wider class of correlations than in the case of standard Bell inequalities.  We conclude that one can find situations in which HV models are in conflict with the conservation of angular momentum law and cannot be represented by states of a physical spin.

\section{Methods}

\subsection{Magnitude conservation for arbitrary $s$}

We look for a solution to the following equation:
\begin{eqnarray}
s_x^2+s_y^2+s_z^2=s(s+1),
\label{EQ:s_eq}
\end{eqnarray}
where we used a simplified notation $s_x=v(\hat{S}_x)$, $s_y=v(\hat{S}_y)$ and $s_z=v(\hat{S}_z)$. The numbers $s$, $s_i$ ($i=x,y,z$) are either all integers or half-integers. We are going to show that: for the half-integer case Eq.~(\ref{EQ:s_eq}) has a solution iff $2s = 1 \Mod 4$, i.e., $s=\frac{1}{2},\frac{5}{2},\frac{9}{2},\ldots$; for the integer case Eq.~(\ref{EQ:s_eq}) has a solution iff $s$ cannot be written in either of the forms:
\begin{eqnarray}
s&=&4(8i+3),\label{EQ:case1}\\
s&=&16(8i+7)\times4^j,\label{EQ:case2}\\
s&=&4(8i+5)-1,\label{EQ:case3}\\
s&=&16(8i+1)\times4^j-1\label{EQ:case4},
\end{eqnarray}
where $i$ and $j$ are non-negative integers. The first few integer values of $s$ for which there is no solution are $s=12,15,19,44,51,\ldots$

The above can be proven by using the Legendre's three-square theorem of 1798 (see e.g. \cite{book1}). It tells that a non-negative integer $n$ can be written as a sum of three squares of integers iff $n$ is not of the form
\begin{eqnarray}
4^a(8b+7),
\label{EQ:n_form}
\end{eqnarray}
where $a$ and $b$ are non-negative integers.

Half-integer case. We multiply Eq.~(\ref{EQ:s_eq}) by 4 and get $(2s_x)^2+(2s_y)^2+(2s_z)^2=2s(2s+2)$, where $2s, 2s_x, 2s_y$ and $2s_z$ are odd integers.
As the right-hand side (RHS) is odd, we only need to consider the case~(\ref{EQ:n_form}) with $a=0$.
Hence, the RHS can be written as a sum of three squares of integers iff it is not $7 \Mod8$, where $\Mod i$ detones modulo-$i$ operation.
We have two cases: either i) $2s=1 \Mod4$, then $4s=2 \Mod8$, thus $(2s)^2+2(2s)=3 \Mod8$, or ii) $2s=3 \Mod4$, then $4s=6 \Mod8$, thus $(2s)^2+2(2s)=7 \Mod8$. Therefore, in the latter case there is no solution.

In the former case, the RHS can be written as a sum of three squares of integers. What we should prove is that all three numbers are squares of odd integers. It is clear that the square of an odd integer is odd and the square of an even integer is even. Next, the sum of squares of three odd integers is always $3 \Mod8$. This is a straightforward consequence of the fact that $(2k+1)^2 = 1 + 8k(k+1)/2$ and $k(k+1)/2$ is an integer if $k$ is an integer. The sum of squares of two even numbers and an odd number (the only other way to get an odd RHS) can not be $3 \Mod8$, because the square of an even number is either $0 \Mod8$ or $4 \Mod8$, so this way the LHS is either $1 \Mod8$ or $5 \Mod8$. This concludes the proof for the half-integer case.

Integer case. How can $s(s+1)$ take the form in Eq.~(\ref{EQ:n_form})?
Since $s(s+1)$ is even, $a$ in Eq.~(\ref{EQ:n_form}) cannot be $0$.

First let $s$ be even. We can assume $s=4^a q$, where $q$ is an odd integer.
This is because the even factor in the product $s(s+1)$ is $4^a$, which  must be contained in $s$, and $s$ must not contain any more even factor.  Then, $s(s+1)=4^a(4^a q^2+q)$. Hence $(4^a q^2+q) 7 \Mod 8$. From previous proof we know that $q^2=1 \Mod8$ which implies $4^a+q=7 \Mod8$.
If $a=1$ the above condition means $q=3 \Mod8$, while if $a > 1$ the above condition means $q=7 \Mod8$.
The former case corresponds to (\ref{EQ:case1}), while the latter one to (\ref{EQ:case2}).

Now let $s$ be odd. In this case, $s+1=4^a q$, where $q$ is an odd integer for the same reason as in the even case. Then $s(s+1)=(4^a q-1)4^a q=4^a(4^a q^2 - q)$. This is of the form Eq.~(\ref{EQ:n_form}) iff $4^a-q=7 \Mod8$ (see argument above).
If $a=1$, then the above condition means $q=5 \Mod8$, while if $a>1$ the above condition means $q=1 \Mod8$.
The former case corresponds to (\ref{EQ:case3}), while the latter one to (\ref{EQ:case4}).

\subsection{Derivation of the inequality for general $s$}

Consider the following state
\begin{equation}\label{singlet}
|\psi_0\rangle = \frac{1}{\sqrt{2s+1}}\sum_{m=-s}^s (-1)^{s-m}|m\rangle \otimes |-m\rangle.
\end{equation}
This is a singlet state of two spin-s particles. It is maximally entangled and the corresponding total angular momentum is zero. As a result, the state is invariant under rotations generated by $e^{i\mathbf{n}\cdot \mathbf{ S^{(A)}}\theta}\otimes e^{i\mathbf{n}\cdot \mathbf{S^{(B)}}\theta}$, where $\theta$ is the angle of rotation of each spin about the $\mathbf{n}$ axis.

Next, let us consider correlations
\begin{equation}
\langle \psi_0|\hat{S}_z^{(A)}\hat{S}_z^{(B)}|\psi_0\rangle = -\frac{1}{2s+1} \sum_{m=-s}^s m^2.
\end{equation}
The above can be evaluated using the formula for square pyramidal numbers (see e.g. \cite{book2})
\begin{equation}
\sum_{k=1}^n k^2 = \frac{n(n+1)(2n+1)}{6}.
\end{equation}
We get
\begin{equation}
\langle \psi_0|\hat{S}_z^{(A)}\hat{S}_z^{(B)}|\psi_0\rangle = -\frac{s(s+1)}{3}.
\end{equation}

The rotational symmetry implies
\begin{eqnarray}
& & \langle \psi_0|\hat{S}_x^{(A)}\hat{S}_x^{(B)}|\psi_0\rangle = \langle \psi_0|\hat{S}_y^{(A)}\hat{S}_y^{(B)}|\psi_0\rangle \nonumber \\
&=& \langle \psi_0|\hat{S}_z^{(A)}\hat{S}_z^{(B)}|\psi_0\rangle = -\frac{s(s+1)}{3},
\end{eqnarray}
Therefore
\begin{equation}
\sum_{j=x,y,z}\langle \psi_0|\hat{S}_j^{(A)}\hat{S}_j^{(B)}|\psi_0\rangle = \langle \mathbf{S^{(A)}} \cdot \mathbf{C_{id}} \cdot \mathbf{S^{(B)}} \rangle_{\psi_0} =-s(s+1),
\end{equation}
where $\mathbf{C_{id}}$ is a $3\times 3$ identity matrix.

Next, consider an Euler rotation of the second spin generated by $\hat U^{(B)} = e^{i \hat{S}^{(B)}_z \theta}e^{i \hat{S}^{(B)}_y \varphi}e^{i \hat{S}^{(B)}_z \xi}$. If this rotation is applied to the singlet state one gets
\begin{equation}
|\phi\rangle = \hat U^{(B)}|\psi_0\rangle.
\end{equation}
As a result
\begin{equation}
\sum_{j=x,y,z}\langle \phi|\hat{S}_j^{(A)}\hat U^{(B)}\hat{S}_j^{(B)}\hat U^{(B)\dagger}|\phi\rangle = -s(s+1).
\end{equation}
However, the above can be rewritten as
\begin{equation}
\langle \mathbf{S^{(A)}} \cdot \mathbf{C} \cdot \mathbf{S^{(B)}} \rangle_{\phi} =-s(s+1),
\end{equation}
where $\mathbf{C}$ is an orthogonal matrix whose entries ($c_{xx}$, $c_{xy}$, $\dots$) are given by
\begin{equation}
\hat U^{(B)}\hat{S}_j^{(B)}\hat U^{(B)\dagger} = c_{jx}\hat{S}_x^{(B)} + c_{jy}\hat{S}_y^{(B)} + c_{jz}\hat{S}_z^{(B)}.
\end{equation}

The other way around. For a pair of arbitrary s-spin particles the operator
\begin{equation}\label{operator}
\mathbf{S^{(A)}} \cdot \mathbf{C} \cdot \mathbf{S^{(B)}},
\end{equation}
where $\mathbf{C}$ is a rotation matrix, has an eigenvalue $-s(s+1)$ with the corresponding eigenvector $|\phi\rangle$. In fact, this is the smallest eigenvalue of the operator (\ref{operator}), because it has the same spectrum as $\mathbf{S^{(A)}} \cdot \mathbf{S^{(B)}}$, for which $-s(s+1)$ is the smallest eigenvalue. This stems from
\begin{equation}
\langle(\mathbf{S^{(A)}}+\mathbf{S^{(B)}})^2 \rangle = \langle (\mathbf{S^{(A)}})^2\rangle + \langle (\mathbf{S^{(B)}})^2\rangle + 2 \langle \mathbf{S^{(A)}}\cdot \mathbf{S^{(B)}}\rangle \geq 0,
\end{equation}
and since  $\langle (\mathbf{S^{(A)}})^2 \rangle = \langle (\mathbf{S^{(B)}})^2 \rangle = s(s+1)$ one gets
\begin{equation}
\langle \mathbf{S^{(A)}}\cdot \mathbf{S^{(B)}}\rangle \geq -s(s+1).
\end{equation}

We know that if $\mathbf{C}$ is a rotation matrix, then in quantum theory the following expression
\begin{equation}
\langle \mathbf{S^{(A)}} \cdot \mathbf{C} \cdot \mathbf{S^{(B)}} \rangle,
\end{equation}
has a lower bound $\beta_Q=-s(s+1)$. In particular, there exists a state $|\phi\rangle$ which corresponds to an eigenvalue $-s(s+1)$. The goal is to show that the classical bound $\beta$ (taking into account the conservation of spin vector magnitude) obeys $\beta > -s(s+1)$.

In case of our HV model the above expression takes form
\begin{equation}
\langle \mathbf{s}^{(A)} \cdot \mathbf{C} \cdot \mathbf{s}^{(B)} \rangle \geq \beta,
\end{equation}
where the average is taken with respect to some probability distribution over vectors $\mathbf{s}^{(X)}=(v(\hat{S}_x^{(X)}),v(\hat{S}_y^{(X)}),v(\hat{S}_z^{(X)}))$ whose entries are integers or half-integers. Moreover, $\mathbf{S}^{(X)}\cdot \mathbf{S}^{(X)} = s(s+1)$, due to the fact that the HV model conserves the spin magnitude.

The lower bound is determined by two vectors $\mathbf{a}$ and  $\mathbf{b}$ which minimize the expression $\mathbf{a}\cdot\mathbf{C}\cdot\mathbf{b}$, i.e.,
\begin{equation}
\beta = \min_{\mathbf{a},\mathbf{b}} \left( \mathbf{a}\cdot\mathbf{C}\cdot\mathbf{b} \right).
\end{equation}
Because $\mathbf{C}$ is a rotation matrix one can write $\mathbf{C}\cdot\mathbf{b}=\mathbf{b}'$, where $\mathbf{b}'\cdot \mathbf{b}' = s(s+1)$. It is in principle possible that $\mathbf{b}' = -\mathbf{a}$, which would lead to $\beta =-s(s+1)$. However, we need to remember that the vector $\mathbf{a}$ is made of integers or half-integers and therefore only a subset of orthogonal matrices of measure zero gives $\mathbf{b}'$ that is also made of integers or half-integers. In particular, for a rotation matrix $\mathbf{C}$ whose entries are irrational numbers one always gets $\beta > -s(s+1)$.

{\bf Data Availability Statement}. Data sharing not applicable to this article as no datasets were generated or  analysed during the current study.

{\bf Acknowledgments}. P.K, W.L and A.K acknowledge support from National Science Center (NCN, Poland) Grant No. 2014/14/M/ST2/00818 and 2016/23/G/ST2/04273. K.F.P and T.V are supported by the National Research, Development and Innovation Office NKFIH (Grant Nos. K111734, and KH125096). J.R acknowledges the National Research Foundation, Prime Minister’s Office, Singapore and the Ministry of Education, Singapore under the Research Centres of Excellence programme.

{\bf Author contributions statements}. The research was initialized by P.K. and W.L. All authors (P.K., W.L., A.K., K.P, J.R. and T.V.) carried out the theoretical calculations and contributed to the disscusions and interpretation of the result. P.K. prepared the manuscript.

{\bf Competing interests}. The authors declare no competing interests.

\end{document}